\documentclass[conference]{IEEEtran}
\IEEEoverridecommandlockouts
\usepackage{cite}
\usepackage{amsmath,amssymb,amsfonts}
\usepackage{algorithmic}
\usepackage{graphicx}
\usepackage{textcomp}
\usepackage{xcolor}
\usepackage{multirow}
\usepackage{url}
\usepackage{array}
\usepackage{booktabs}
\usepackage{pifont}
\usepackage{graphicx}
\newcommand{\cmark}{\ding{51}}
\newcommand{\xmark}{\ding{55}}

\def\BibTeX{{\rm B\kern-.05em{\sc i\kern-.025em b}\kern-.08em
    T\kern-.1667em\lower.7ex\hbox{E}\kern-.125emX}}

\begin{document}

\title{Disentangling and Fusing Neurostructural and Vascular Ageing for Retinal Age Prediction}

% ===== AUTHORS: commented out for double-blind review =====
% Restore before camera-ready submission.
%
\author{%
\begin{minipage}{\textwidth}
\centering

% ---------- First row ----------
\begin{minipage}[t]{0.24\textwidth}
\centering
{\normalsize 1\textsuperscript{st} Junwen Zheng}\\
{\small
\textit{College of Computing}\\
\textit{and Data Science}\\
\textit{Nanyang Technological}\\
\textit{University}\\
Singapore\\
JUNWEN003@e.ntu.edu.sg
}
\end{minipage}\hfill
\begin{minipage}[t]{0.24\textwidth}
\centering
{\normalsize 2\textsuperscript{nd} Li Rong Wang}\\
{\small
\textit{College of Computing}\\
\textit{and Data Science}\\
\textit{Nanyang Technological}\\
\textit{University}\\
Singapore\\[0.4em]
\textit{Centre for Frontier AI}\\
\textit{Research, A*STAR}\\
Singapore\\
LIRONG002@e.ntu.edu.sg
}
\end{minipage}\hfill
\begin{minipage}[t]{0.24\textwidth}
\centering
{\normalsize 3\textsuperscript{rd} Anthony Zihan Lin}\\
{\small
\textit{College of Computing}\\
\textit{and Data Science}\\
\textit{Nanyang Technological}\\
\textit{University}\\
Singapore\\
anth0029@e.ntu.edu.sg
}
\end{minipage}\hfill
\begin{minipage}[t]{0.24\textwidth}
\centering
{\normalsize 4\textsuperscript{th} Xinran Xu}\\
{\small
\textit{Lee Kong Chian}\\
\textit{School of Medicine}\\
\textit{Nanyang Technological}\\
\textit{University}\\
Singapore\\
XINRAN007@e.ntu.edu.sg
}
\end{minipage}

\par\vspace{1.5em}

% ---------- Second row ----------
\begin{minipage}[t]{0.24\textwidth}
\centering
{\normalsize 5\textsuperscript{th} Wei Kiong Ngo}\\
{\small
\textit{Tan Tock Seng Hospital}\\
Singapore\\
wei\_kiong\_ngo@ttsh.com.sg
}
\end{minipage}\hfill
\begin{minipage}[t]{0.24\textwidth}
\centering
{\normalsize 6\textsuperscript{th} Kelvin Zhenghao Li}\\
{\small
\textit{Centre of AI in Medicine}\\
\textit{Lee Kong Chian}\\
\textit{School of Medicine}\\
\textit{Nanyang Technological}\\
\textit{University}\\
Singapore\\[0.4em]
\textit{National Healthcare}\\
\textit{Group Eye Institute}\\
Singapore\\
kelvin.lizh@ntu.edu.sg
}
\end{minipage}\hfill
\begin{minipage}[t]{0.24\textwidth}
\centering
{\normalsize 7\textsuperscript{th} Tock Han Lim}\\
{\small
\textit{Lee Kong Chian}\\
\textit{School of Medicine}\\
\textit{Nanyang Technological}\\
\textit{University}\\
\textit{National Healthcare Group}\\
Singapore\\
tock.han.lim@nhghealth.com.sg
}
\end{minipage}\hfill
\begin{minipage}[t]{0.24\textwidth}
\centering
{\normalsize 8\textsuperscript{th} Xiuyi Fan\textsuperscript{*}}\\
{\small
\textit{Lee Kong Chian}\\
\textit{School of Medicine and}\\
\textit{College of Computing}\\
\textit{and Data Science}\\
\textit{Nanyang Technological}\\
\textit{University}\\
Singapore\\
xyfan@ntu.edu.sg
}
\end{minipage}

\end{minipage}%
\thanks{*Corresponding author: Xiuyi Fan (email: xyfan@ntu.edu.sg).}
}
% \author{\IEEEauthorblockN{Anonymous Submission}}
% ==========================================================

\maketitle

\begin{abstract}
Estimating biological age is an important task in ageing research, as it quantifies individual ageing trajectories beyond chronological age. Retinal age estimation has become a well-established direction in this area because retinal imaging provides a non-invasive window into neural and microvascular ageing. Existing studies, however, have predominantly focused on fundus photographs and usually model retinal ageing as a single generic process. Although biological ageing is heterogeneous and different organs or tissues may age at different rates, little research has explicitly distinguished neurostructural and vascular ageing in retinal age prediction.
This work fills this gap by studying neurostructural ageing with Optical Coherence Tomography (OCT) images and vascular ageing with Optical Coherence Tomography Angiography (OCTA) images. We formulate multimodal OCT/OCTA retinal age estimation as a structural--vascular ageing decomposition problem and propose SAP-DPF, a unified dual-path prediction framework that separately estimates structural and vascular ageing, models modality-specific predictive uncertainty, and adaptively fuses the two ageing signals through an uncertainty-gated late-fusion module. Using this unified prediction framework, SAP-DPF achieves a mean absolute error of 4.07 years for the structural pathway and 7.37 years for the vascular pathway, while the proposed late-fusion algorithm further improves the overall mean absolute error to 4.02 years, representing a 12.52\% improvement over the strongest single-modality baseline. This work could extend retinal age prediction beyond a single biological-age estimate, providing a framework for investigating structural and vascular contributions to heterogeneous retinal ageing.

\end{abstract}


%%% ============================================================
\section{Introduction}
%%% ============================================================

Retinal ageing is a clinically meaningful biological ageing phenotype because the retina captures both neural and microvascular changes through non-invasive imaging.  The retinal age gap, defined as the difference between image-predicted retinal age and chronological age, has been associated with all-cause mortality~\cite{zhu2023retinal}, cardiovascular disease~\cite{nusinovici2022retinal}, and incident dementia~\cite{sim2025deep}. These findings suggest that retinal age may reflect systemic ageing trajectories.

Retinal ageing has been most extensively studied using color fundus photography~\cite{zhu2023retinal,nusinovici2022retinal,ahadi2023longitudinal,yu2025cross,nielsen2024foundation,ninomiya2026high}. However, fundus photography provides a projected view in which neural tissue appearance, optic-disc morphology, pigmentation, and vascular geometry are jointly encoded. As a result, most fundus-based retinal age predictors produce a single retinal-age estimate and do not explicitly model whether ageing-related information arises from neurostructural change, vascular alteration, or their combination. Depth-resolved OCT and vascular OCTA offer a complementary opportunity to study these ageing components more directly, but OCT/OCTA-based retinal age prediction remains comparatively underexplored.
\begin{figure*}[t]
\centering
\includegraphics[trim={0.2cm 0.2cm 0.2cm 0.2cm},clip,width=0.82\textwidth]{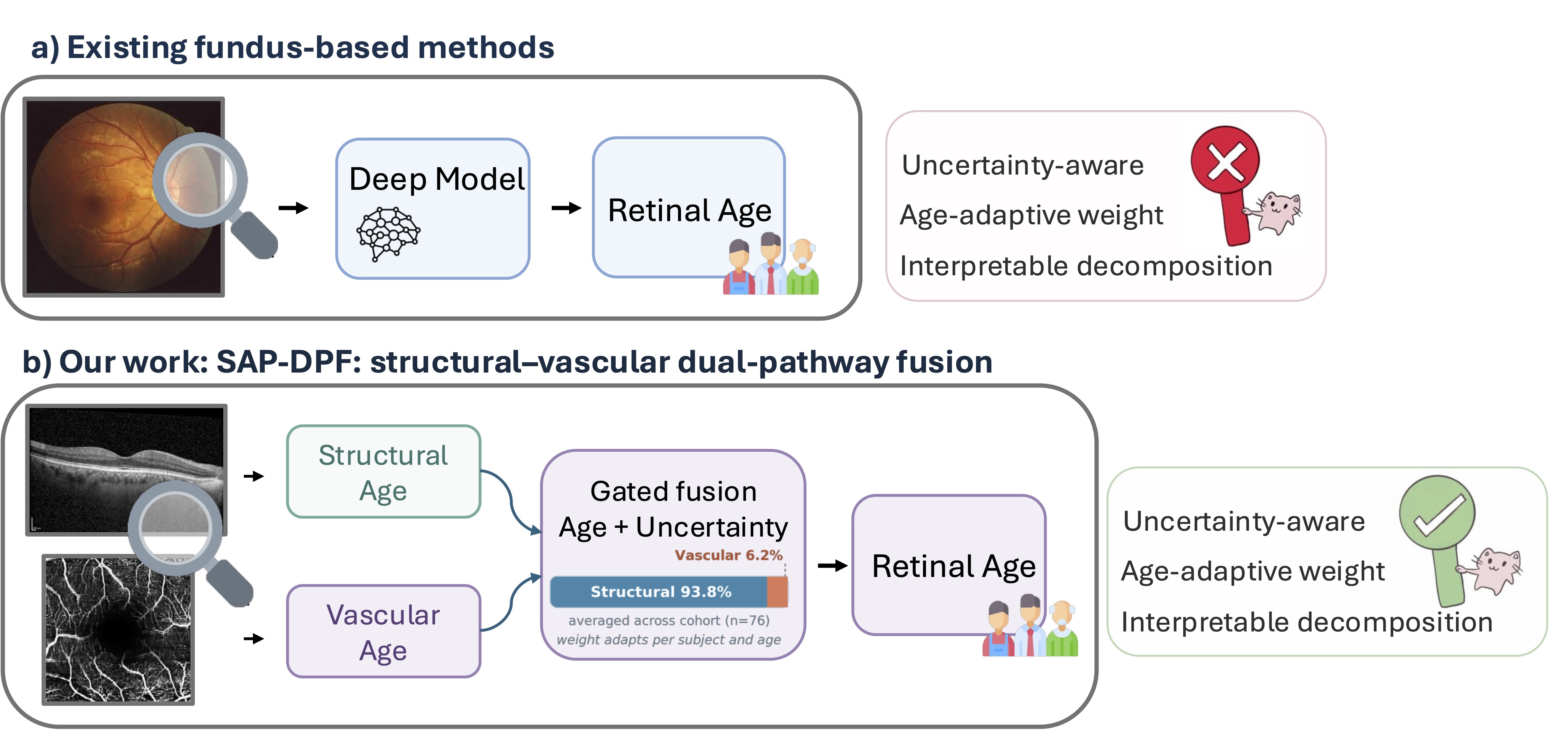}
\caption{\textbf{Comparison between fundus-based retinal age prediction and SAP-DPF.}
\textbf{(a)} Existing fundus-based methods usually produce a single overall retinal age estimate from fundus photographs.
\textbf{(b)} SAP-DPF uses OCT and OCTA to estimate neurostructural age, vascular age, and a reliability-aware fused retinal age.}
\label{fig:comparison}
\end{figure*}

\begin{table*}[!t]
\centering
\caption{Qualitative comparison of fusion strategies for multimodal
OCT--OCTA age estimation, complementing the quantitative results in
Table~\ref{tab:results}. Column definitions: \textbf{Uncertainty-Aware},
per-sample predictive variance (Eq.~\eqref{eq:ensemble_var});
\textbf{Age-Adaptive}, reliability varies smoothly with age
(Eq.~\eqref{eq:age_prior}); \textbf{Subject Gate}, per-sample convex
gate over the two pathways (Eq.~\eqref{eq:gate}); \textbf{Dual-Path},
structural/vascular groups fused before inter-group fusion (cf.\
\textit{w/o Dual-Path}, Table~\ref{tab:results}); \textbf{Traceable
Weights}, weight plots as an explicit function of age/pathway
(Figs.~\ref{fig:intra_weights},~\ref{fig:pi_scatter},~\ref{fig:pi_age}).}
\label{tab:qualitative}
\renewcommand{\arraystretch}{1.3}
\setlength{\tabcolsep}{5pt}
\resizebox{\textwidth}{!}{%
\begin{tabular}{@{}
  >{\raggedright\arraybackslash}m{3.3cm}
  >{\centering\arraybackslash}m{1.9cm}
  >{\centering\arraybackslash}m{1.6cm}
  >{\centering\arraybackslash}m{1.7cm}
  >{\centering\arraybackslash}m{1.5cm}
  >{\centering\arraybackslash}m{1.9cm}
  >{\raggedright\arraybackslash}m{5.0cm}@{}}
\toprule
\textbf{Method} & \textbf{Uncertainty-Aware} & \textbf{Age-Adaptive} & \textbf{Subject Gate} & \textbf{Dual-Path} & \textbf{Traceable Weights} & \textbf{Key Limitation} \\
\midrule
Simple Average & \xmark & \xmark & \xmark & \xmark & \xmark & Assumes uniform modality reliability across all subjects and ages. \\
\midrule
IVW (MAE)~\cite{sherratt2024characterising} & \xmark & \xmark & \xmark & \xmark & \xmark & Static, population-level weight from validation-set error; fixed regardless of subject or age. \\
\midrule
IVW (Uncertainty)~\cite{lee2023development} & \cmark & \xmark & \xmark & \xmark & \xmark & Sample-level but age-invariant; weight reflects predictive variance only, with no pathway-level meaning. \\
\midrule
Learned Weights~\cite{couvy2020ensemble} & \xmark & \xmark & \xmark & \xmark & \xmark & A single global scalar per modality; not resolved by subject or age. \\
\midrule
Stacking (Ridge/ElasticNet)~\cite{wolpert1992stacked,le2021deep} & \xmark & \xmark & \xmark & \xmark & \xmark & Strong point accuracy but global meta-learner coefficients; uncertainties, if used, enter only as meta-features. \\
\midrule
\textbf{SAP-DPF (Ours)} & \cmark & \cmark & \cmark & \cmark & \cmark & --- \\
\bottomrule
\end{tabular}
}
\end{table*}

OCT and OCTA provide distinct ophthalmic measurements that are well suited to neurostructural--vascular decomposition. OCT offers depth-resolved neurostructural imaging of retinal layer morphology, including macular and circumpapillary anatomy and retinal nerve fibre layer features~\cite{hormel2023oct}. Age-related changes in retinal nerve fibre layer and choroidal measures have been quantified in OCT studies~\cite{celebi2013age,zhou2020age}. OCTA provides non-invasive visualization of retinal and choroidal microvasculature, enabling assessment of vascular plexuses and choriocapillaris patterns relevant to systemic and ocular disease~\cite{chua2024optical}; OCT- and OCTA-derived retinal features have also been associated with visual function in older adults~\cite{dong2022association}. At the same time, OCTA measurements may be more vulnerable to acquisition artifacts, segmentation variability, and longitudinal reproducibility limits~\cite{nishida2023long}. These modality differences suggest that neurostructural and vascular ageing should be estimated separately and fused according to their reliability, rather than treated as interchangeable multimodal inputs.

We propose Smooth Age-Prior Dual-Path Fusion (SAP-DPF), a multimodal OCT/OCTA framework for retinal age estimation. SAP-DPF treats retinal ageing as three related outputs rather than a single fused prediction: a neurostructural ageing pathway estimated from OCT, a vascular ageing pathway estimated from OCTA, and a reliability-aware fusion of the two pathways. This design reflects the medical premise that neurostructural and vascular ageing may encode distinct biological information and may differ in reliability across subjects and ages. Because vascular ageing estimates can be noisier and more confounded than neurostructural estimates, the final fusion is designed to be reliability-aware, age-adaptive, and subject-specific. SAP-DPF uses modality-specific age regressors, predictive uncertainty estimates, smooth age-dependent reliability priors, and a gated late-fusion mechanism to integrate the two ageing pathways without obscuring their individual contributions. Fig.~\ref{fig:comparison} illustrates this distinction: existing fundus-based approaches generally yield one overall retinal age estimate, whereas SAP-DPF derives neurostructural age, vascular age, and a reliability-aware fused retinal age from OCT/OCTA.

\begin{figure*}[t]
\centering
\includegraphics[width=0.95\textwidth]{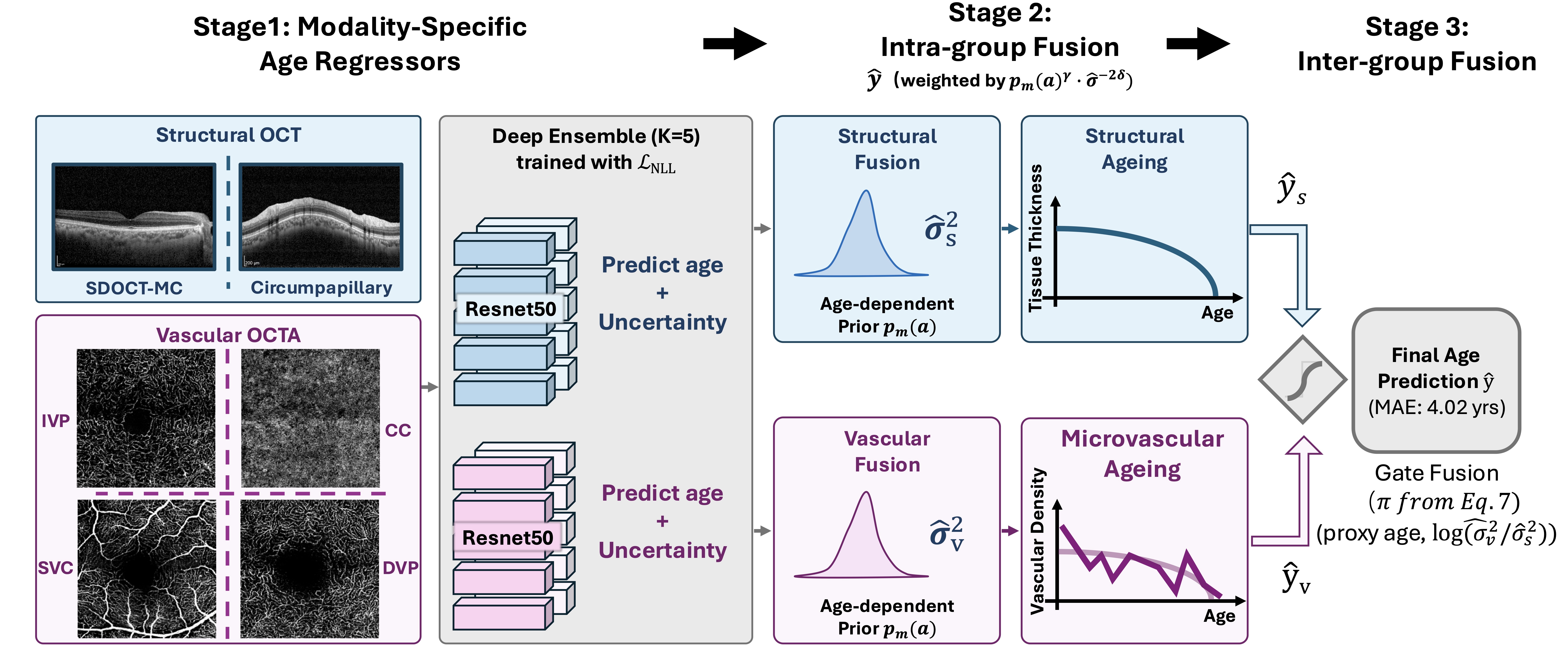}
\caption{\textbf{The SAP-DPF framework.}
\textbf{Stage~1:} Six modalities are partitioned into structural (SDOCT-MC, CP) and vascular (IVP, CC, SVC, DVP) groups. Each modality is processed using a deep ensemble ($K{=}5$) of ResNet-50 regressors, producing age predictions with predictive variances.
\textbf{Stage~2:} Within each group, predictions are fused via reliability-aware weighted averaging, combining smooth modality priors $p_m(a)$ with sample-level inverse variance. This yields group-level estimates $\hat{y}_S$ and $\hat{y}_V$ for the structural and vascular pathways, respectively.
\textbf{Stage~3:} A soft gate $\pi$ (Eq.~\eqref{eq:gate}), parameterised by the proxy age and the log variance ratio, integrates the two pathway estimates into the final prediction $\hat{y}$.}
\label{fig:overview}
\end{figure*}

Our main contributions are: (1) We formulate multimodal OCT--OCTA age estimation as a structural--vascular ageing decomposition problem and instantiate it with SAP-DPF, a dual-path framework that fuses the two pathways according to predictive reliability. (2) We introduce smooth reliability priors to characterise modality-specific accuracy variation across age, together with a gated mechanism that balances the two pathways using individual-level prediction confidence. (3) On a six-modality OCT--OCTA dataset, SAP-DPF achieves an MAE of 4.02 years, representing a 12.52\% improvement over the strongest single-modality baseline, while the learned fusion weights provide interpretable insight into age-dependent structural and vascular contributions.

%%% ============================================================
\section{Background}
%%% ============================================================

\subsection{Age Prediction from Retinal Imaging}
Existing deep learning approaches to retinal age prediction have predominantly relied on single imaging modalities. Early work focused on color fundus photographs, where predicted retinal age gaps were associated with mortality, cardiovascular, and cognitive
outcomes~\cite{zhu2023retinal,nusinovici2022retinal, ahadi2023longitudinal,nielsen2024foundation}. Subsequent studies moved to depth-resolved OCT, first predicting age from peripapillary~\cite{shigueoka2021predicting} and later macular scans~\cite{chen2024deep}, progressively exploiting richer anatomical
detail for age estimation. These efforts confirm that both fundus and OCT imaging carry reliable age-informative signals, but each estimates retinal age from a single modality and does not exploit the complementarity between structural and vascular imaging. 

\subsection{Fusion Algorithms for Multimodal Prediction}

Once two complementary modalities are in hand, retinal age estimation
turns into a question of how to combine them, and it is here that
existing designs fall short. The few multimodal ageing studies to date
have relied on simple combination rules. Le Goallec et
al.~\cite{le2021deep} paired fundus and OCT predictors and observed
that the two carry partially non-overlapping ageing signals, yet
merged them with a fixed stacked combiner. Closest to our OCT/OCTA
setting, Pourjavan et al.~\cite{pourjavan2025ai} regressed ocular age
from predefined structural and vascular metrics, offering early
evidence of structural--vascular complementarity but drawing on
hand-crafted measurements rather than representations learned from the
images themselves. These works compute the weights for each modality once and apply that weight to every subject. In practice, however, the reliability of a retinal scan is neither fixed nor shared across a cohort. Motion artifacts or signal attenuation can degrade one subject's OCT image while leaving the same modality accurate for others; a modality that predicts age well in younger patients may not generalise to older ones, as anatomical ageing cues shift over the lifespan. Because reliability varies by both subject and age, a single pre-set weight cannot track either effect. Hence, retinal scan reliability needs to be estimated per subject and age.
% Previous version: These schemes weight each modality once and apply that weight to everyone. In practice, however, the reliability of a retinal scan is neither fixed nor shared across a cohort. Motion artifacts or signal attenuation can degrade one subject's image while leaving the same modality accurate for others, and a modality that predicts age well in younger eyes need not do so in older ones, as anatomical ageing cues shift over the lifespan. A single pre-set weight cannot track either effect; doing so requires reliability to be estimated per subject and per age.

Existing decision-level fusion strategies approach this requirement to varying degrees, but none fully achieves it. The simplest, population-level schemes—plain averaging or validation-set inverse-variance weighting~\cite{sherratt2024characterising}—assign one weight per modality and reuse it for every subject. Uncertainty-based inverse-variance weighting~\cite{lee2023development} takes the first step toward adaptivity, letting per-sample predictive variance modulate the weights, but the resulting weights remain age-invariant and carry no notion of a structural or vascular pathway. Learned and stacked combiners~\cite{couvy2020ensemble,wolpert1992stacked,le2021deep} fit modality weights from data and often reach strong point accuracy, yet their coefficients are global, fixed once training ends, and difficult to interpret clinically. Gated
fusion~\cite{arevalo2017gated} enables sample-conditioned weighting, but has not been tied to age or to the structural--vascular split. What no prior strategy offers is a weight that is at once uncertainty-aware, age-adaptive, subject-specific, and attributable to a structural or vascular pathway. Table~\ref{tab:qualitative} makes this gap explicit, and it is the gap SAP-DPF is built to close.

%%% ============================================================
\section{Methods}
%%% ============================================================

We employ a unified late-fusion framework for retinal age prediction that jointly supports structural age estimation, vascular age estimation, and their adaptive integration. Here, structural age and vascular age denote the age estimates derived from the OCT and OCTA pathways, respectively. Rather than treating OCT and OCTA as generic multimodal inputs, the framework maintains neurostructural and vascular ageing as two explicit prediction pathways before combining them at the decision level. This design yields separate pathway-level age estimates, allows the final prediction to be examined through pathway-specific contributions, and limits the influence of less reliable vascular signals on the overall prediction. 
\subsection{Modality-Specific Age Regressors}
\label{sec:modality_regressors}

We consider six imaging modalities: two structural OCT scans---circumpapillary (CP) and Single Horizontal B-Scan of Spectral Domain Optical Coherence Tomogram across the Macular Centre (SDOCT-MC)---and four OCTA en face slabs: Superficial Vascular Complex (SVC), Intermediate Vascular Plexus (IVP), Deep Vascular Plexus (DVP), and Choriocapillaris (CC). The modality set is denoted as
\[
\mathcal{M} = \{\text{SDOCT-MC},\, \text{CP},\, \text{SVC},\, \text{IVP},\, \text{DVP},\, \text{CC}\}.
\]

Since retinal image quality varies across modalities and subjects due to factors such as motion artifacts and signal attenuation, each modality carries irreducible observation noise (aleatoric uncertainty); for each modality $m \in \mathcal{M}$, an independent regressor $f_{\theta_m}$ takes an input image $x_{i,m}$ and produces both an age estimate and an aleatoric uncertainty estimate:
\begin{equation}
\bigl[\,\hat{\mu}_{i,m},\; \log \hat{\sigma}_{i,m}^{2}\bigr]
= f_{\theta_m}\!\left(x_{i,m}\right),
\end{equation}
where $\hat{\mu}_{i,m}$ denotes the predicted age and $\hat{\sigma}_{i,m}^{2}$ represents the sample-level heteroscedastic variance. Each regressor employs a ResNet-50 backbone initialised with ImageNet-pretrained weights, followed by two parallel linear heads that output the mean and log-variance, respectively.

To enable reliability-aware fusion, each regressor explicitly estimates heteroscedastic aleatoric uncertainty, capturing modality- and sample-specific noise characteristics. Following~\cite{kendall2017uncertainties}, the regressors are trained using the Gaussian negative log-likelihood loss:
\begin{equation}
\mathcal{L}_{\mathrm{NLL},m}
= \frac{1}{N}\sum_{i=1}^{N}
\left[
\frac{\left(y_i - \hat{\mu}_{i,m}\right)^{2}}
     {2\,\hat{\sigma}_{i,m}^{2}}
+ \frac{1}{2}\log \hat{\sigma}_{i,m}^{2}
\right],
\end{equation}
where $N$ is the number of training samples and $y_i$ denotes the chronological age of sample $i$.

To complement the per-regressor aleatoric estimates, we employ a deep ensemble of $K$ independently trained regressors per modality to capture epistemic uncertainty, which reflects model uncertainty due to limited training data~\cite{kendall2017uncertainties,lakshminarayanan2017simple}.  Given sample $i$ and modality $m$, the ensemble predictive mean and total variance are:
\begin{align}
  \bar{\mu}_{i,m} &= \frac{1}{K}\sum_{k=1}^{K}\hat{\mu}_{i,m}^{(k)}, \label{eq:ensemble_var} \\
  \bar{\sigma}_{i,m}^{2}
  &= \underbrace{\frac{1}{K}\sum_{k=1}^{K}\hat{\sigma}_{i,m}^{2,(k)}}_{\text{aleatoric}}
  + \underbrace{\frac{1}{K}\sum_{k=1}^{K}
      \bigl(\hat{\mu}_{i,m}^{(k)} - \bar{\mu}_{i,m}\bigr)^{2}}_{\text{epistemic}}. \nonumber
\end{align}

\subsection{Intra-group Fusion: Reliability-aware Fusion within OCT and OCTA}
\label{sec:intra_fusion}
We adopt a decision-level (late) fusion strategy, aggregating each modality's age prediction and uncertainty rather than fusing intermediate features across modalities. This choice reflects the asymmetric noise characteristics of OCT and OCTA: fusing features before per-modality reliability can be assessed risks propagating vascular noise into a shared representation and would make it difficult to disentangle each pathway's contribution to the final prediction.

Since structural and vascular modalities exhibit different ageing dynamics, we first aggregate predictions within each modality group before merging the two pathways. We partition $\mathcal{M}$ into structural $\mathcal{S}=\{\text{SDOCT-MC, CP}\}$ and vascular $\mathcal{V}=\{\text{SVC, DVP, IVP, CC}\}$, fusing predictions within each group first.

\paragraph{Age-dependent reliability prior}
Modality informativeness varies with age, e.g., CP is more discriminative in older subjects while SDOCT-MC carries stronger signal in younger ones. Because modality reliability is expected to vary smoothly with age, we define $p_m(a)$ for each modality~$m$ as the reciprocal of a kernel-smoothed estimate of local prediction error, following the Nadaraya--Watson estimator~\cite{nadaraya1964estimating}:
\begin{align}
    p_m(a) &= \left(
        \frac{\sum_{j} \kappa_h(a - a_j)\,\bigl|a_j - \bar{\mu}_{j,m}\bigr|}
             {\sum_{j} \kappa_h(a - a_j)}
        + \epsilon\right)^{-1}\!, \label{eq:age_prior} \\
    \kappa_h(u) &= \exp\!\Bigl(-\frac{u^{2}}{2h^{2}}\Bigr), \nonumber
\end{align}
where $a_j$ is the chronological age of validation sample~$j$, $\bar{\mu}_{j,m}$ the corresponding ensemble prediction from modality~$m$, $h$ is a kernel bandwidth selected via leave-one-out cross-validation, and $\epsilon$ is a small constant for numerical stability.

\paragraph{Intra-group fusion}
For a sample~$i$, the fused prediction within $\mathcal{G} \in \{\mathcal{S}, \mathcal{V}\}$ follows a reliability-weighted scheme~\cite{baltruvsaitis2018multimodal}:
\begin{align}
  \hat{y}_{i,\mathcal{G}}
  &= \sum_{m\in\mathcal{G}} w_{i,m}\,\bar{\mu}_{i,m}, \label{eq:intra_fusion} \\
  w_{i,m}
  &\propto
  \bigl[p_m(\tilde{a}_i)\bigr]^{\gamma}
  \cdot
  \bigl[\bar{\sigma}_{i,m}^{-2}\bigr]^{\delta}, \nonumber
\end{align}
where $\tilde{a}_i$ is a \emph{proxy age} obtained by averaging the structural modality predictions (thus available at test time without ground-truth labels), $\bar{\sigma}_{i,m}^{2}$ is the total predictive variance defined in Eq.~\eqref{eq:ensemble_var} (combining aleatoric and epistemic uncertainty across ensemble members), and $\gamma,\delta > 0$ are learnable exponents. Together, $p_m(\tilde{a}_i)$ captures population-level age-varying reliability, while $\bar{\sigma}_{i,m}^{-2}$ down-weights modalities that are unreliable for a particular subject. The fused group-level variance is propagated as
\begin{equation}
\label{eq:group_var}
\bar{\sigma}_{i,\mathcal{G}}^{2} = \sum_{m \in \mathcal{G}} w_{i,m}^{2}\,\bar{\sigma}_{i,m}^{2}.
\end{equation}

\subsection{Inter-group Fusion with Gated Mechanism}

To integrate structural and vascular estimates, we employ an adaptive fusion strategy in which the final prediction is a convex combination of the two pathway estimates:
\begin{align}
\hat{y}_i &= \pi_i \hat{y}_{i,\mathcal{S}} + (1 - \pi_i)\hat{y}_{i,\mathcal{V}}, \label{eq:gate} \\
\pi_i &= \sigma \!\left(
k_0 + k_1 \bar{a}_i + k_2 \bar{a}_i^2
+ k_3 \log \frac{\bar{\sigma}^2_{i,\mathcal{V}}}{\bar{\sigma}^2_{i,\mathcal{S}}}
\right), \nonumber
\end{align}
where $\sigma(\cdot)$ denotes the sigmoid function, $\bar{a}_i$ is the standardized proxy age, and $\bar{\sigma}^2_{i,\mathcal{S}}$ and $\bar{\sigma}^2_{i,\mathcal{V}}$ are the structural and vascular group-level uncertainties, respectively, obtained from Eq.~\eqref{eq:group_var}.

In Eq.~\eqref{eq:gate}, we adopt a low-order polynomial function with parameters $k_0$, $k_1$ and $k_2$ to model the slowly varying relationship between biological ageing and modality relevance. Retinal structural and vascular alterations evolve non-linearly yet gradually~\cite{su2022age,trinh2021modelling}; a second-order polynomial captures broad curvature effects while limiting model complexity. The sigmoid formulation ensures bounded, continuous weighting, yielding a stable convex combination of modality-specific predictions. The log-variance ratio term introduces uncertainty-aware modulation, allowing the gate to adjust the pathway weights according to their relative predictive uncertainty.

%%% ============================================================
\section{Experiment}
%%% ============================================================

\subsection{Dataset}
We conducted experiments on a dataset derived from the Health for Life in Singapore (HELIOS) study~\cite{wang2025health}, a population-based prospective cohort study. Subjects underwent deep phenotyping including multimodal retinal imaging. Subjects were proportionately sampled across both genders and 5-year age cohorts (30 to $\geq$80 years). Exclusions included incomplete imaging and high myopia ($\geq$6 diopters). The subset contains 504 subjects with chronological age labels and multi-modal retinal scans in six imaging modalities of two eyes. Following standard quality control, we excluded scans with missing modalities and severe image-quality issues. This results in 1,008 images per image modality. To ensure statistical independence, all splits were performed at the subject level. The final dataset was partitioned into 352 subjects (704 images per modality) for training, 76 subjects (152 images per modality) each for validation and testing. The mean age was 55.4 years (SD 14.7; range 30.2--83.2).

\begin{table}[htbp]
\caption{Single-modality and intra-group fusion results on the held-out test set ($n{=}152$ images, 76 subjects). $\Delta$MAE: improvement over the best single modality in each group (Circumpapillary for Structural, DVP for Vascular).}
\label{tab:hierarchical}
\begin{center}
\setlength{\tabcolsep}{4pt}
{\fontsize{8pt}{9pt}\selectfont
\begin{tabular}{llcccc}
\hline
Group & Modality & MAE$\downarrow$ & RMSE$\downarrow$ & $R^2$$\uparrow$ & $\Delta$MAE \\
\hline
\multirow{3}{*}{Structural}
 & SDOCT-MC        & 4.71 & 5.83  & .846 & --- \\
 & Circumpapillary & 4.59 & 5.73  & .852 & --- \\
 & \textit{Structural Fusion} & \textbf{4.07} & \textbf{5.05} & \textbf{.885} & $-$0.52 \\
\hline
\multirow{5}{*}{Vascular}
 & CC  & 9.51 & 11.28 & .425 & --- \\
 & SVC & 9.68 & 12.12 & .336 & --- \\
 & IVP & 8.26 & 10.66 & .486 & --- \\
 & DVP & 7.54 & 10.29 & .522 & --- \\
 & \textit{Vascular Fusion} & \textbf{7.37} & \textbf{10.01} & \textbf{.547} & $-$0.17 \\
\hline
\end{tabular}
}
\end{center}
\end{table}

\begin{table}[htbp]
\caption{Fusion strategy comparison (upper) and ablation study (lower)
on the test set ($n{=}152$ images, 76 subjects). \textbf{Bold} denotes
best result. IVW: Inverse Variance Weighting. $\Delta$MAE: difference
relative to SAP-DPF (positive $=$ worse). NLL: Gaussian negative
log-likelihood on the test set (lower indicates better Gaussian
probabilistic predictive performance). Subject MAE is computed by averaging
predictions across both eyes per subject. Ablation variants: \textit{w/o
Gate} fixes $\pi{=}0.5$; \textit{w/o Dual-Path} fuses all six modalities
in a single group; \textit{w/o Smooth Prior} replaces kernel-smoothed
age-adaptive priors with uniform priors.}

\label{tab:results}
\begin{center}
\setlength{\tabcolsep}{2pt}
{\fontsize{7.5pt}{9pt}\selectfont
\begin{tabular}{lccccccc}
\hline
Method & MAE$\downarrow$ & Subj.MAE$\downarrow$ & RMSE$\downarrow$ & $R^2$$\uparrow$ & $r$$\uparrow$ & $\Delta$MAE & NLL$\downarrow$ \\
\hline
\multicolumn{8}{l}{\textit{Fusion strategy comparison}} \\
Simple Average          & 5.98 & 5.83 & 7.74 & .729 & .886 & $+$1.96 & 4.382 \\
IVW (MAE)~\cite{sherratt2024characterising}        & 5.43 & 5.27 & 6.96 & .781 & .911 & $+$1.41 & 4.214 \\
IVW (Uncertainty)~\cite{lee2023development} & 4.26 & 4.15 & 5.27 & .875 & .941 & $+$0.24 & 3.853 \\
Learned Weights~\cite{couvy2020ensemble}   & 4.22 & 4.00 & 5.23 & .876 & .940 & $+$0.21 & 3.987 \\
Stacking (Ridge)~\cite{wolpert1992stacked}  & 4.23 & 4.03 & 5.27 & .874 & .936 & $+$0.21 & 3.082 \\
Stacking (ElasticNet)~\cite{le2021deep}    & 4.15 & 3.96 & 5.19 & .878 & .939 & $+$0.14 & 3.061 \\
\textbf{SAP-DPF (Ours)} & \textbf{4.02} & \textbf{3.79} & \textbf{5.01} & \textbf{.887} & \textbf{.942} & --- & \textbf{3.036} \\
\hline
\multicolumn{8}{l}{\textit{Ablation study}} \\
w/o Gate ($\pi{=}0.5$)  & 5.01 & --- & 6.39 & .815 & .906 & $+$0.99 & --- \\
w/o Dual-Path           & 4.10 & --- & 5.09 & .883 & .940 & $+$0.08 & --- \\
w/o Smooth Prior        & 4.06 & --- & 5.03 & .886 & .941 & $+$0.05 & --- \\
\hline
\end{tabular}
}
\end{center}
\end{table}

\begin{figure}[htbp]
\centering
\includegraphics[width=\linewidth]{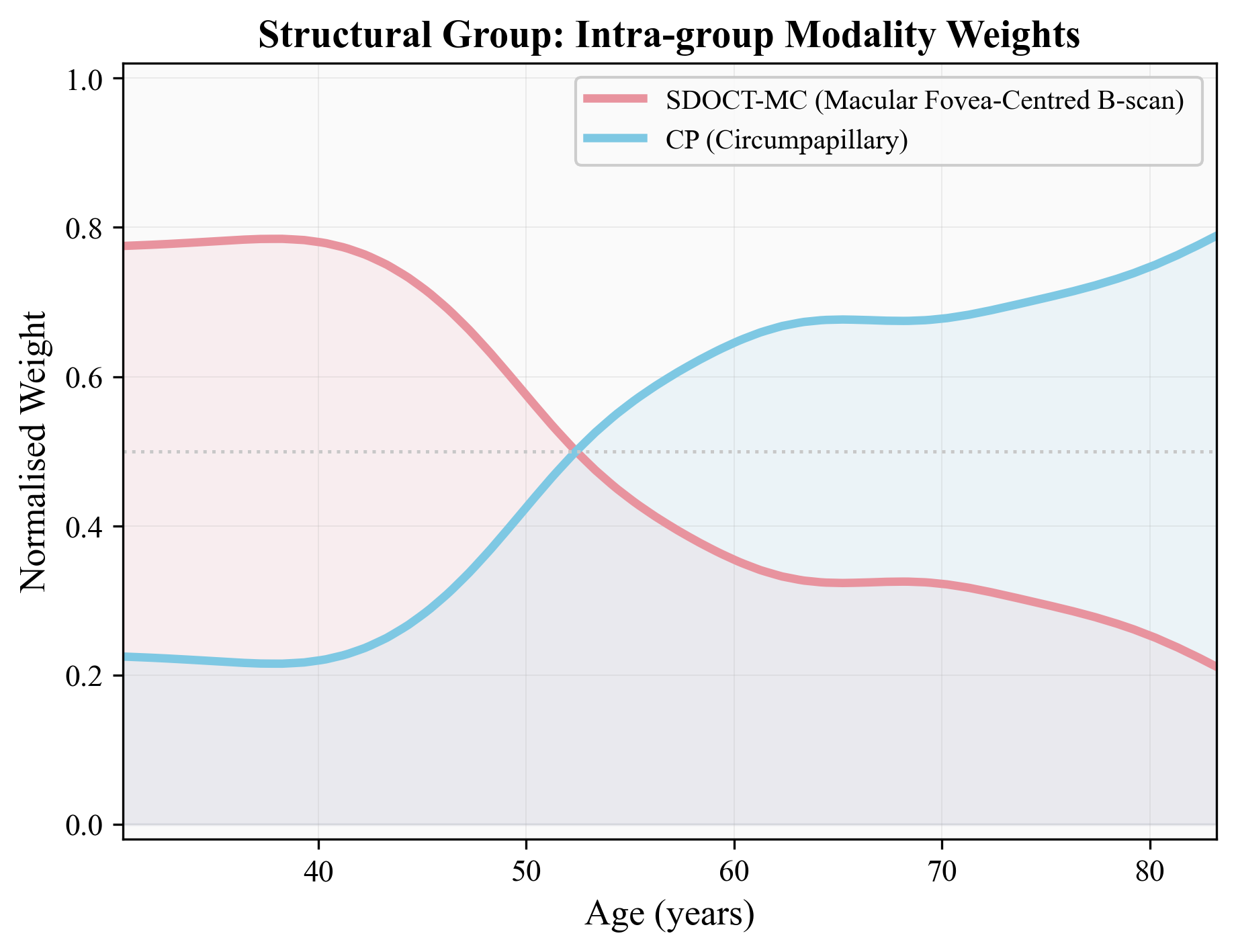}
\caption{Intra-group fusion weights for structural modalities (SDOCT-MC and CP) derived from smooth reliability priors $p_m(a)$ (Eq.~\eqref{eq:age_prior}). SDOCT-MC dominates in younger subjects while CP becomes dominant beyond age~$\approx$50, consistent with age-dependent circumpapillary changes.}
\label{fig:intra_weights}
\end{figure}

\begin{figure}[htbp]
\centering
\includegraphics[width=\linewidth]{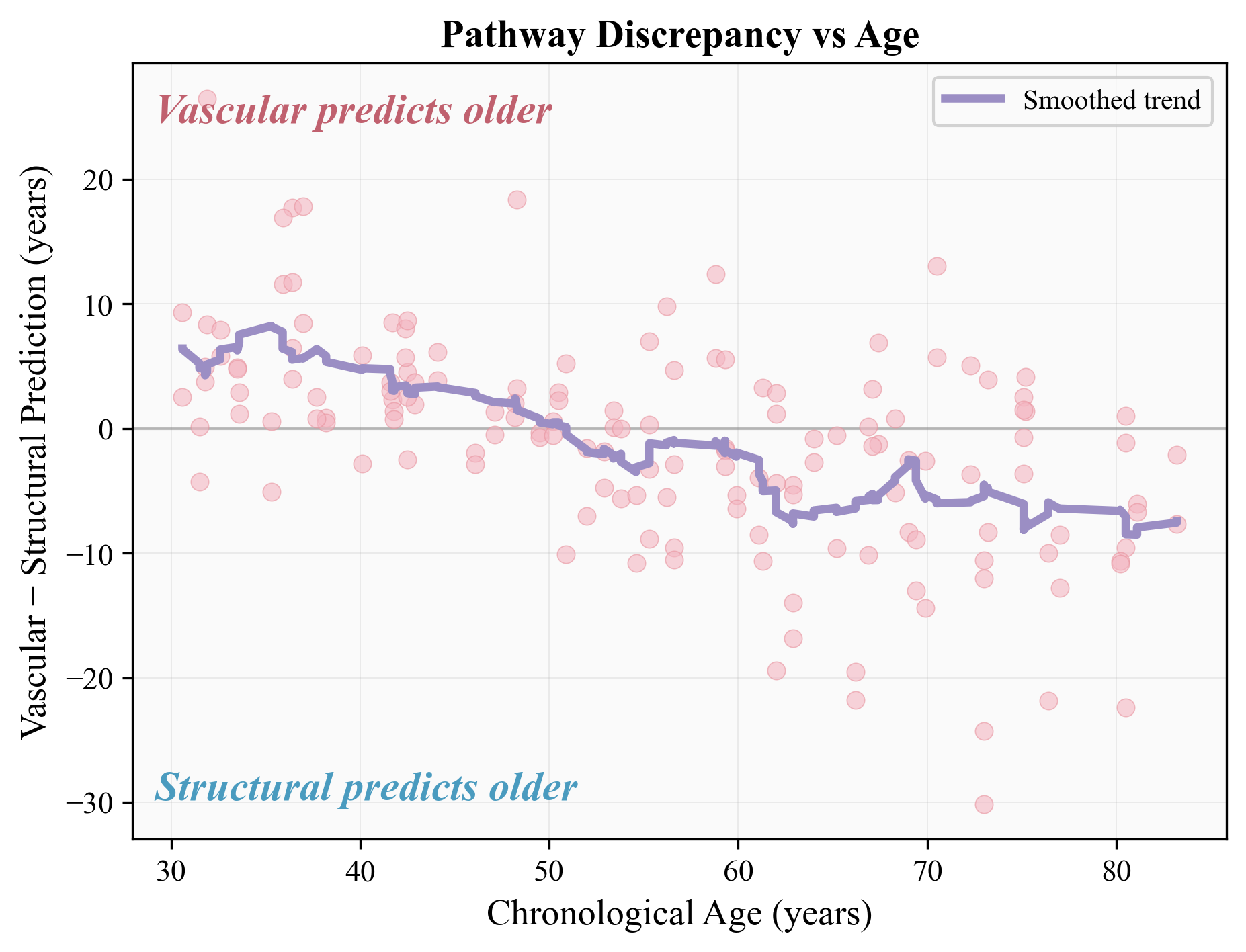}
\caption{Prediction discrepancy ($\hat{y}_V - \hat{y}_S$) against chronological age. The shift from positive discrepancy in younger individuals to negative discrepancy in older individuals highlights age-dependent differences between structural and vascular age estimates.}
\label{fig:intra_discrepancy}
\end{figure}

\begin{figure}[htbp]
\centering
\includegraphics[width=\linewidth]{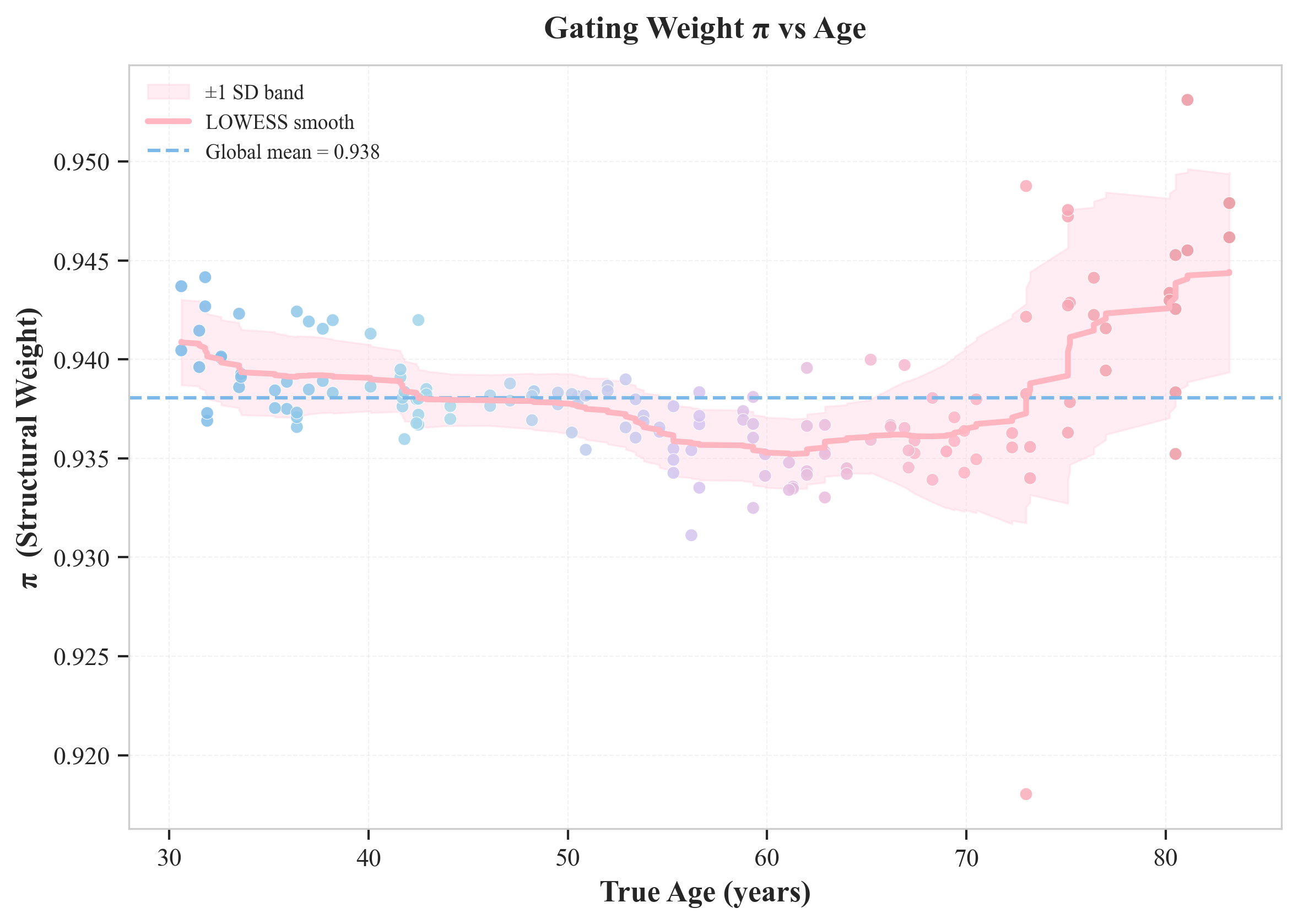}
\caption{Structural gating weight $\pi$ against chronological age (LOWESS smooth $\pm$1\,SD; dashed line: global mean $\pi{=}0.938$). $\pi$ shows a modest dip in the 50--70 age range, suggesting a slight relative increase in vascular contribution at these ages.}
\label{fig:pi_scatter}
\end{figure}

\begin{figure}[htbp]
\centering
\includegraphics[width=\linewidth]{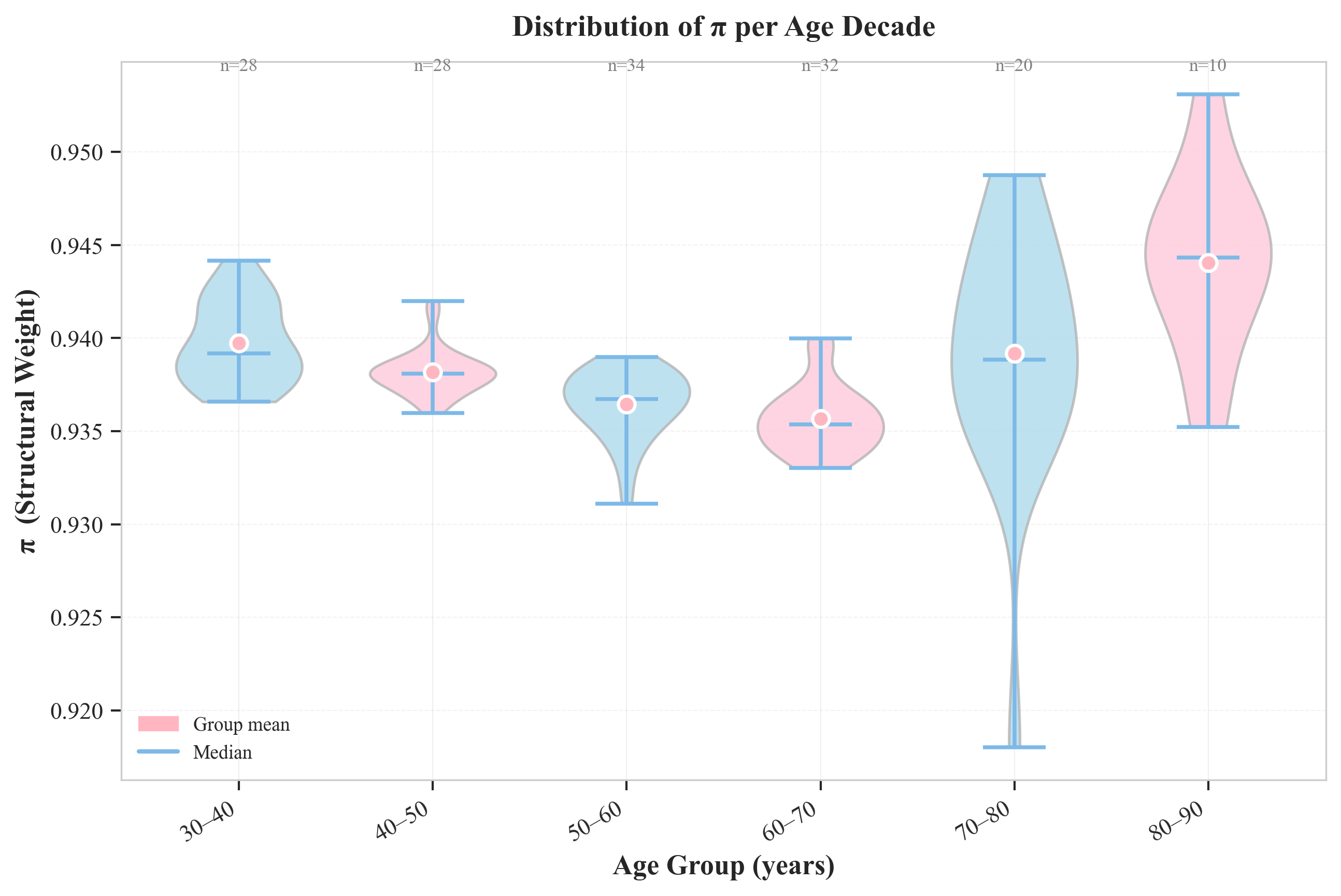}
\caption{Distribution of structural gating weight $\pi$ per age decade. The widening violin in the 70--80 and 80+ decades suggests increased variability in structural--vascular balance at older ages.}
\label{fig:pi_age}
\end{figure}

\subsection{Implementation Details}
\label{sec:implementation}
All modality-specific regressors use a ResNet-50 backbone pretrained on ImageNet. Each modality trains a deep ensemble of $K\!=\!5$ members via Adam ($lr\!=\!10^{-3}$, batch size 32, early stopping patience 25). Fusion hyperparameters (kernel bandwidth $h$, gate regularisation weight $\lambda$) are selected via leave-one-out cross-validation; remaining fusion parameters are optimised with differential evolution on the validation set. The remaining parameters, $\gamma$, $\delta$, and the gating coefficients $k_0, \dots, k_3$, are optimised on the validation set by minimising mean absolute error, with $\ell_2$ regularisation (weight $\lambda$) applied to $k_1, k_2, k_3$ to discourage overfitting of the age-dependent gate. Modality uncertainties are rescaled by a learned monotone transformation before weighting, and the fused output is further refined by a bias correction and a variance calibration, all fitted on the validation set.

\subsection{Prediction Performance}

Structural OCT yields lower single-modality error than vascular OCTA, and intra-group fusion further reduces error (Table~\ref{tab:hierarchical}). SAP-DPF achieves the lowest overall MAE among evaluated fusion strategies (MAE$=$4.02 yr, $R^2{=}0.887$; Table~\ref{tab:results}) with an average structural weight $\pi = 0.938$, indicating predominantly structural predictions with a modest vascular refinement, while the vascular pathway additionally provides a separate vascular age estimate. SAP-DPF significantly outperforms simple fusion baselines (Simple Average, IVW-MAE~\cite{sherratt2024characterising}, IVW-Uncertainty~\cite{lee2023development}; subject-level Wilcoxon signed-rank test, $n{=}76$; $p \leq 0.021$), with stacking methods performing comparably ($\Delta$MAE$\leq$0.21).  %new 
% Table~\ref{tab:qualitative} summarises this distinction qualitatively,
% showing that SAP-DPF is the only strategy satisfying all five criteria
% (see caption for definitions).

%new
Beyond point-prediction accuracy, we evaluated probabilistic predictive performance using Gaussian negative log-likelihood (NLL), which jointly depends on the predicted mean and variance (Table~\ref{tab:results}, rightmost column). SAP-DPF achieved the lowest observed NLL among the fusion methods. Wilcoxon signed-rank tests on subject-averaged NLL ($n{=}76$) showed significant NLL improvements over Simple Average, IVW-MAE, IVW-Uncertainty, and Learned Weights (all $p<0.0001$). Against the strongest stacking baselines, SAP-DPF showed small NLL differences, with nominal significance versus Stacking Ridge ($p=0.040$) and no significant difference versus ElasticNet ($p=0.088$). This indicates
that SAP-DPF achieves probabilistic performance comparable to strong stacking-based fusion while offering a more interpretable uncertainty-aware fusion mechanism. 
Results hold at the subject level (Subject MAE$=$3.79 vs.\ 3.96 for the strongest stacking baseline; Table~\ref{tab:results}).

As a further robustness check, we performed subject-level 5-fold cross-validation over the pooled 152 validation and test subjects, refitting all six Table~\ref{tab:results} baselines and SAP-DPF per fold while keeping the modality regressors fixed. SAP-DPF won 5/5, 5/5, and 4/5 folds against the baselines it significantly outperforms in single-split
point-prediction MAE (Simple Average, IVW-MAE, and IVW-Uncertainty), and 4/5, 3/5, and 3/5 folds against the baselines with which it is statistically tied (Learned Weights, Stacking Ridge, and ElasticNet), respectively. This pattern is consistent with Table~\ref{tab:results} and supports that the single-split comparison is not an artifact of one particular partition. Together with the NLL results, this suggests that SAP-DPF's added value relative to well-tuned stacking baselines lies in interpretable, uncertainty-aware fusion with competitive probabilistic performance, rather than uniformly superior point-prediction accuracy.
%new
To assess the local sensitivity of the validation-set-selected fusion parameters, we perturbed $\gamma$, $\delta$, $h$, and $\lambda$ by $\pm10\%/\pm20\%$ around their selected values, and the gate coefficients $k_0$--$k_3$ individually, re-evaluating on the held-out test set without re-selecting the other hyperparameters (for $h$ and $\lambda$, the remaining fusion parameters were refitted on the validation set). Test MAE varied by less than 0.1 years for every parameter (Table~\ref{tab:sensitivity}), including all four gate coefficients ($k_0$: 0.050y, $k_1$: 0.0001y, $k_2$: 0.002y, $k_3$: 0.007y), indicating the fusion stage is not sensitive to small deviations from the selected values. 

\begin{table}[t]
\centering
\caption{Learned fusion hyperparameters and sensitivity to
$\pm10\%/\pm20\%$ perturbation (max$-$min test-set MAE, years).}
\label{tab:sensitivity}
\begin{tabular}{lcc}
\hline
Parameter & Learned value & MAE range \\
\hline
$\gamma$ (age-prior exponent)   & 3.05  & 0.024 \\
$\delta$ (precision exponent)   & 0.46  & 0.009 \\
$h$ (bandwidth, yr)             & 5     & 0.097 \\
$\lambda$ (gate $\ell_2$ reg.)  & 0.5   & 0.094 \\
$k_0$ (gate bias)               & 3.00  & 0.050 \\
$k_1$ (age-linear)              & $-0.01$ & 0.0001 \\
$k_2$ (age-quadratic)           & 0.11  & 0.002 \\
$k_3$ (vasc-uncertainty)        & 0.12  & 0.007 \\
\hline
\end{tabular}
\end{table}

In eyes of elderly subjects ($\geq$70 years) with above-median OCTA uncertainty ($n{=}16$ eyes), SAP-DPF reduces MAE by 5\% over ElasticNet~\cite{le2021deep}, the strongest stacking baseline (4.576 vs.\ 4.820), though confirmation on larger cohorts is warranted. Ablations show the largest drop when removing the gate, and the explicit structural--vascular decomposition yields interpretable fusion weights.

The predicted uncertainties are empirically validated: Gaussian 95\% prediction intervals covered 94.1\% of test eyes (143/152), and modality-level aleatoric uncertainty correlates positively with prediction error for the four OCTA modalities (Pearson $r$ up to 0.34, Spearman $\rho$ up to 0.44), consistent with the rationale for inverse-variance weighting in Eq.~\eqref{eq:intra_fusion}.

\subsection{Clinical Interpretation}

Our framework provides interpretable signals that characterise \emph{how} structural and vascular substrates contribute to retinal ageing estimation. Within the structural pathway, the learned reliability prior shifts from favouring macular OCT in younger subjects to circumpapillary OCT in older subjects, with a crossover near age 50 (Fig.~\ref{fig:intra_weights}). This pattern indicates age-dependent differences in predictive reliability between the two structural modalities in this cohort. At the inter-group level, the structural pathway remains the primary driver across the cohort ($\pi = 0.938$), while a modest reduction in midlife (50--70\,years; Fig.~\ref{fig:pi_scatter}) and increased variability in older subjects (Fig.~\ref{fig:pi_age}) indicate variation in the vascular pathway's contribution to the fused prediction. The two pathways exhibit a systematic discrepancy that transitions from positive (vascular predicts older) in younger subjects to negative in older subjects (Fig.~\ref{fig:intra_discrepancy}), suggesting different age-dependent prediction patterns across structural and vascular imaging. This structural--vascular discordance motivates further investigation of heterogeneous retinal ageing. Structural OCT provided the more accurate age estimate, whereas vascular age variability may reflect both imaging
limitations~\cite{nishida2023long} and systemic cardiovascular risk factors. Future studies should assess whether accounting for these factors improves chronological age prediction, while also investigating whether the vascular age gap provides complementary information about cardiovascular risk.

\section{Conclusion}

We presented SAP-DPF, a structural--vascular dual-path framework for multimodal OCT--OCTA retinal age prediction. The framework separately estimates structural age from OCT and vascular age from OCTA before integrating the two predictions. Structural OCT provided the more accurate age estimate, while vascular predictions offered a modest complementary  improvement, yielding a fused MAE of 4.02 years.

SAP-DPF combines smooth age-dependent reliability priors and sample-level predictive uncertainty for intra-group fusion, followed by an uncertainty-aware gate for subject-specific integration. This design preserves pathway-specific age estimates and interpretable fusion weights, enabling analysis of structural--vascular prediction differences alongside overall predictive performance. These outputs provide a basis for investigating heterogeneous retinal ageing, although their biological significance requires further validation.

The present study is limited to a single cohort of moderate size, and generalisation across populations, imaging devices, and acquisition protocols remains to be established. The vascular pathway also remained substantially less accurate. Future studies should assess whether accounting for systemic cardiovascular risk factors improves age prediction, and whether pathway-specific age gaps provide information about cardiovascular and other clinical outcomes beyond established risk factors. Such analyses should account for age-related prediction bias and image quality and include independent longitudinal validation.

\section*{Acknowledgment}

This research is supported by the Singapore Ministry of Health's
National Medical Research Council under its Population Health
Research Grant MOH-002116-00 and the Centre of AI in Medicine (C-AIM).

\bibliographystyle{IEEEtran}
\bibliography{references}

\end{document}